\documentclass[graybox]{svmult}

\usepackage{mathptmx}       
\usepackage{helvet}         
\usepackage{courier}        
\usepackage{type1cm}        
\usepackage{makeidx}         
\usepackage{graphicx}        
\usepackage{multicol}        
\usepackage[bottom]{footmisc}

\makeindex             

\begin{document}

\title*{IoT Lotto: Utilizing IoT Devices in Brute-Force Attacks}
\author{Mohammed M. Alani}
\authorrunning{M.M. Alani}
\institute{Mohammed M. Alani \at Khawarizmi International College, Abu Dhabi, United Arab Emirates, \email{m@alani.me}}
%
%

\maketitle
\footnote{This is a pre-print version of the paper. The original paper is published by the ACM as part of the proceedings of the 2018 The 6th International Conference on Information Technology: IoT and Smart City (ICIT 2018).}
\abstract{The number of IoT devices in use is increasing rapidly and so is the number of IoT applications. As in any new technology, the rapid development means rapid increase in security threats and attack surfaces.\\
IoT security has proven to be challenging throughout the past few years. However, another challenging task is to prevent IoT devices from becoming a tool used by malicious attackers to break into other systems.\\
In this paper, we present a conceptual design in which IoT devices are used as tools in brute-force attacks to break encryption keys of block ciphers. The proposed design shows that with adequate number of IoT devices employed in the attack, the attack can succeed in breaking large-key block ciphers.}

\section{Introduction}
As identified in \cite{ref2}, the term Internet-of-Things (IoT) refers to the networked interconnection of everyday objects,
which are often equipped with ubiquitous intelligence. These objects can be your home objects, like the lights, refrigerator, and television, or they can be in a complex industrial environment like intelligent robots in a factory.\\
Growing from around 15 billion devices in 2015, the number of IoT devices connected to the Internet around the world have exceeded 23 billion in 2018 \cite{ref1}. Being an essential building block in Industry 4.0, IoT adoption rates are expected to grow at higher rates in the near future. Ref \cite{ref1} expects the number of IoT devices to grow to over 75 billion in 2025. This rapid growth year-over-year brings new security challenges.\\
When IoT devices started gaining popularity, the main challenge for mobile IoT devices was battery life\cite{ref5}. Mobile IoT devices were mostly composed of sensor(s), wireless communication component, along with the processing unit. At a later stage, battery issues were mostly resolved with the development of low-power component that can operate on very low power.\\
Common security threats studies in information security are mostly focused on data leakage or service availability. However, threats to IoT are considered mroe serious because they can directly pause physical risk\cite{ref6}. Commonly used IoT devices incorporate sensors and some kind of actuator. In addition these devices communicate with each other. Hence, compromised devices at home or in a factory can cause a lot of damage.\\
Another category of IoT devices is wearable devices. Any compromise in the security of these devices can cause leakage to highly private data. These wearable devices commonly contain multiple types of sensors that are design to collect large amounts of data. Wearable devices have a wide range of applications in areas like fitness tracking and vital signs monitoring.\\
Common security challenges in IoT are roughly summarized in the following non-exhaustive list:
\begin{enumerate}
\item Object identification and authentication.\\
Authentication and object identification can be a challenging task in the IoT environment because of the lack of a centralized authentication entity\cite{ref7}.
\item Data protection.\\
A reasonable countermeasure is to provide end-to-end encryption. However, not all IoT devices have the resources needed for end-to-end encryption like processing power and memory availability\cite{ref8}. In addition, data protection is required while the data is stored in IoT device, not only in transit.
\item Threats to availability.\\
The availability of IoT devices can be targeted using Denial-of-Service (DoS) and Distributed-Denial-of-Service (DDoS) attacks.
\item Unauthorized access\cite{ref8}.\\
IoT devices, when unauthorized access is gained, can easily be controlled by the malicious attacker to perform tasks that the devices should not be performing.
\item Man-in-the-middle (MITM) attacks.\\
Given that IoT devices rely on wireless communications, MITM attacks have a better chance of succeeding, especially when combined with the lack of a centralized authentication entity.
\item Compatibility threats.\\
Conventional network security protocols were built to serve users exchanging data, and they were not designed to serve machines communicating with each other\cite{ref8}.
\item Application threats.\\
Being a new and rapidly developing area, application software written for IoT is expected to have security weaknesses and loopholes that can cause many zero-day attacks. Patching will be particularly more complicated as IoT devices are designed to work with small memory and storage availability. Hence, software updates are nto as easy as they seem in personal computers, or networking devices.
\end{enumerate}
The reader can easily note that the list above seems similar to security challenges in any computer-based system and not necessarily IoT. However, the mechanism in which these threats are exploited, the counter-measures needed, and even the impact of exploiting these threats differ significantly when IoT is involved. For example, DDoS attack carried out on a wearable IoT device would not only render the device unusable during the attack, rather it would consume the battery power and forces the device to shut off even after the end of the attack.\\

\section{Block-ciphers and Brute-Force Attacks}
In cryptographic terminology, the text before encryption is referred to as plaintext, while the text after encryption is named ciphertext.\\
Block ciphers are encryption algorithms that are design to operate on data in the form of blocks. A block of data, or plaintext, is sent into the algorithm with a key. The algorithm manipulates the bits of the plaintext to produce the ciphertext. This manipulation happens based on the value of the key. One of the earliest standards of block ciphers was the Data Encryption Standard (DES)\cite{ref10}. DES uses a key size of 56-bits and generates sub-keys that are used in the encryption process that goes for 16 rounds to ensure more secure ciphertext. As mentioned in \cite{ref10}, DES was officially adopted as a standard in 1976. At that time, the computing power publicly available was not capable of carrying out a brute-force attack against a 56-bit key.\\
Brute-force attack is a basic attack that searches for the right encryption key by trying out all possible keys. Although this type of attack was thought of as impossible on the DES 56-bit key when it was first adopted, many years later, the DES encryption algorithm was not secure anymore. In addition to the huge increase in computing power and the growth in the possibility of success of brute-force attacks, other attacks were capable of reducing the search space or retrieving parts of the key. Attacks like differential cryptanalysis and linear cryptanalysis managed to reduce the search space from the $2^{56}$ \cite{ref11,ref12}.\\
A newer standard, Advanced Encryption Standard (AES), was adopted in 2001\cite{ref13}. The newer standard was more future oriented and adopted key sizes of 128, 192, and 256 bits. It was also designed to counter the attacks that were possible on DES.\\
Both AES and DES can be categorized as symmetric ciphers. A symmetric cipher is an encryption algorithm that employs the same key for encryption and decryption. The other category; asymmetric ciphers are encryption algorithm that use a key for encryption and a different key for decryption. The most popular asymmetric cipher is RSA\cite{ref14}. In this paper, we will focus on symmetric ciphers brute-force attack.

\section{IoT Device Hijacking}
Before the popularity of IoT, node hijacking was not uncommon. Many security flaws were detected over the years that can be exploited to gain control over a node. The node can be a personal computer, a server, or a network-connected resource. One of the common aims of hijacking a node, is to use the node to perform an attack on another node. Hackers have software crawlers that scavenger the Internet looking for servers or end-user computers with weak protection measures. Once the malicious attacker gains control over the node, the node can be used as a platform to perform attacks on other targets. A common attack that employs hacked nodes is Distributed Denial-of-Service (DDoS) attack. In a DDoS attack, the malicious attacker orchestrates a massive flood of requests directed towards a server or a cluster of servers to render its service unavailable to its rightful users.\\
One of the earliest wide-scale IoT attacks took place in 2014 when malicious attackers broke into over 100,000 home IoT devices like televisions and fridges. The attackers used these hijacked devices to target individuals and businesses around the world with malicious emails\cite{ref16}.\\
In 2016, another DDoS attack was executed using millions of web cameras, printers, and baby monitors. The attack had a stunning bandwidth of 1.2 Tbps, which was the highest at its time. The unlucky target of the attack was an internet services provider and infrastructure operator named Dyn. This attack made many of Dyn's clients' websites unavailable in different parts of the world like Twitter, Paypal, Amazon, CNN, Spotify, and WSJ\cite{ref17}.\\
In 2017, an IoT hijacking botnet named Mirai was discovered. The version infecting windows-based  IoT devices was a result of developments of the malicious software that previously infected personal computers. According to \cite{ref15}, there is currently over half a million devices currently that reside in over 164 countries around the world. In one of its stages of development, Mirai-infected IoT devices jumped from 213000 to 483000 in the period of two weeks.\\
Another famous malware, that infects linux-based IoT devices is BASHLITE (also known as Lizkebab, Torlus, and gafgyt)\cite{ref15}. This malware also uses the hijacked devices to perform DDoS attacks. It is said that this malware has infected over 1 million IoT devices, and can launch DDoS attack of up 400 Gbps capacity.\\
IoT devices were also used in attacks that did not involve compromising or hijacking the IoT device. This type of attacks is referred to as reflective attacks. In a reflective attack, the attacker sends a query (like TCP syn request) with q spoofed source IP address to have all response directed to the target IP address. This was, the attacker does not directly attack the target and also amplifies traffic to inflict more damage\cite{ref16}.\\

\section{Proposed Design}
In 1991, Quisquater and Desmedt introduced the idea of an attacked named the Chinese Lotto attack \cite{ref3}. This attack is based on a simple idea; instead of using supercomputers to launch an exhaustive key-search attack, use a larger number of regular computers\cite{ref4}. The concept introduced relies on using distributed computing to search the complete key space in a relatively short amount of time.\\
The Chinese lotto attack suggests that a massive number of television devices with a decryption capability added to them can be used to decrypt an encrypted message. The massively large key search space in this brute-force attack is divided into smaller spaces. Each one of these small search spaces is then assigned to a specific television set. This way, the search space will be exhausted in a much faster way because it will be searched by all the televisions in China in parallel. The television that finds the right key, would signal a message to the television owner that he/she has won the lotto and they should deliver the key code to the authorities to receive the prize.\\
In the proposed conceptual design we use a similar concept but employing a massive number of IoT devices instead of computers. As brute-force attacks rely on exhaustive search of the whole key space to find the correct decryption key, the proposed system divides the search space into smaller spaces that can be searched in a reasonable amount of time by one IoT device.\\
The proposed design employs a central arbitration unit named the Key Distribution Arbiter (KDA). This arbiter is responsible for assigning keys to IoT devices to assure that each device is assigned an exclusive key space. Figure 1 shows an overview of the proposed design. This arbiter can be hosted on a single server or can be hosted on a cloud solution. Cloud hosting of the arbiter is highly recommended for implementations that include a large number of IoT devices to assure seamless operation of the system.\\
\begin{figure}[htbp]
\includegraphics[scale=.65]{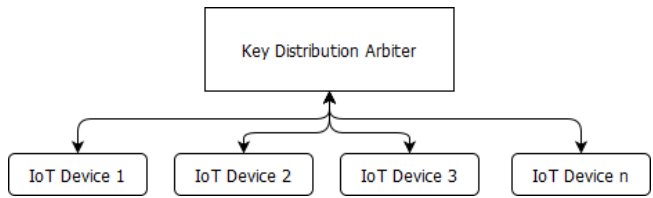}
\caption{Overview of IoT Lotto}
\label{fig1}
\end{figure}
The proposed system uses \textit{n} IoT devices. This arrangement will divide the key search space to \textit{n} smaller search spaces. The main search space has a size of $2^u$ where \textit{u} is the key length of the original symmetric encryption key, measured in bits.\\
Using \textit{n} IoT devices would results in having \textit{n} smaller search spaces each of the size $\frac{2^u}{n}$. This means that the time required to exhaust all of the search space, \textit{t}, will also be divided by \textit	{n} to be $\frac{t}{n}$. The higher \textit{n} goes, the less time it takes to find the right key.\\
Inside each IoT device, three software components are required, as shown in Figure 2. The Key Arbitration Agent (KAA) is the component that coordinates with the KDA to assure that the key search space assigned to the particular device is exclusive. The second component is the decryption algorithm. This component depends on the algorithm that was used in encryption as selected earlier by the attacker. The last component is the one that flags the stop sign when the correct key has been found.\\
\begin{figure}
\includegraphics[scale=0.8]{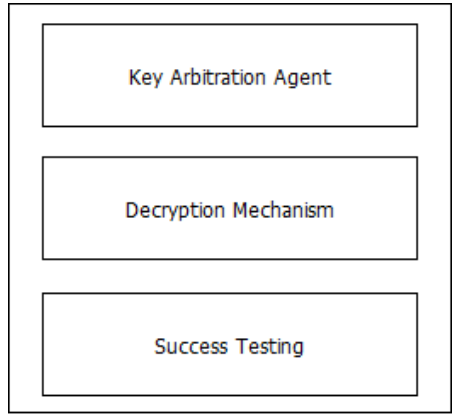}
\caption{Software Components in IoT Devices}
\label{fig2}
\end{figure}
\subsection{Key Arbitration Process}
As mentioned earlier, each IoT device is assigned an exclusive key search space. As the search process is exhaustive and aims at trying to decrypt the given ciphertext block with all possible key combinations, each device will be assigned a range of keys instead of a random key sequence.\\
For example, if $u=4$, then there are $2^4$ possible combinations. Assuming that we have 4 IoT devices, $n=4$, each device is expected to try $\frac{2^4}{4}=4$ keys. Instead of transferring the keys and storing them in the memory of the IoT devices, each device is given the first key and last key in its search space. Hence, for the first IoT device, the first key is 0000, and the last key is 0011. For the second IoT device, the first key is 0100 and the last key is 0111. The keys in-between these ranges, can easily be generated by adding 1 until reaching the last key. Figure 3 shows a flowchart of the key generation process along with the decryption.\\
\begin{figure}
\centering
\includegraphics[scale=0.8]{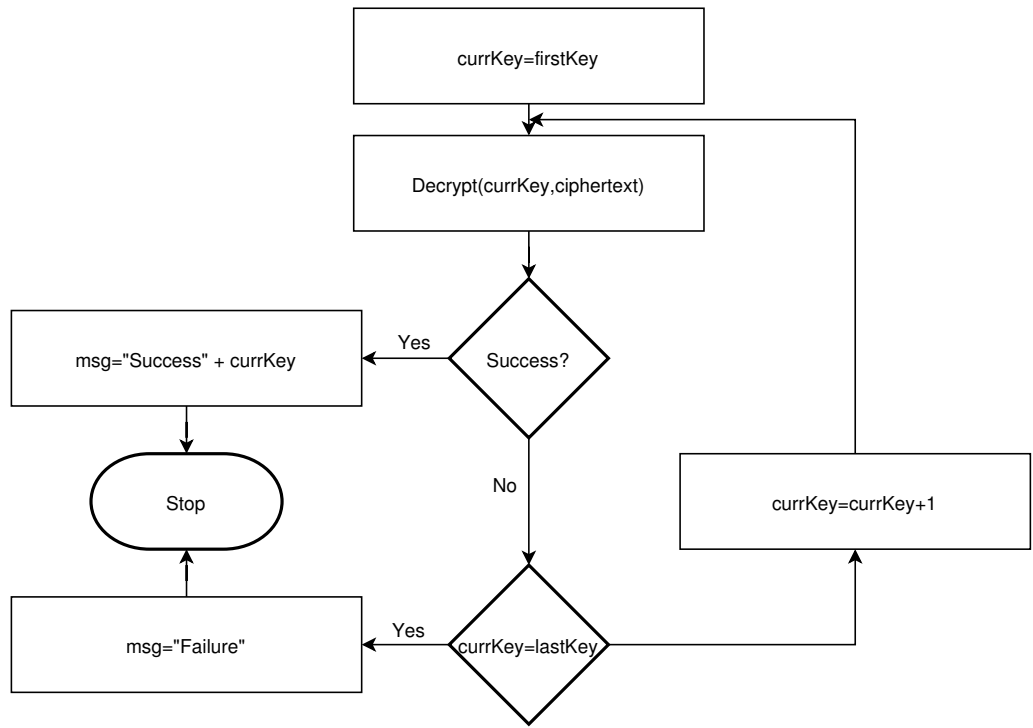}
\caption{Flowchart of key generation and decryption at the IoT device}
\label{fig3}
\end{figure}
This type of key arbitration reduces the memory requirements at the IoT device side and reduces the amount of data to be transferred between the KDA and the IoT device, keeping in mind that the example explained earlier is a non-realistic one because key sizes can be 128, 256, 512, 1024 bits, or higher.
\subsection{Key Search Process}
As shown in Figure 3, after the first and last keys are received from the KDA, the KAA starts generating keys and passing them to the decryption process along with the ciphertext block that is to be decrypted. The process of key generation continues until either success condition is positive or the whole assigned search space is exhausted.
\subsection{Success Condition}
Just like any other brute-force attacks, IoT lotto success condition is achieved when the correct decryption key is identified. This can happen in different ways. One of which is that the ciphertext block used comes from a previously known plaintext block. Another way is to pass the resulting decrypted text into a dictionary comparison to see if it is a word or not. In our conceptual design we adopted the first method which relies on having a known plaintext-ciphertext pair prior to the decryption process.
\section{Discussion}
The proposed conceptual design introduces a system in which a large group of IoT devices can be used as decryption machines to find an encryption key through brute-force attack. The proposed system reduces the time required for the exhaustive search in the key space.\\
The time reduction comes from the fact that the large network of IoT devices are searching in a parallel fashion in different parts of the search space. When the system employs \textit{n} IoT devices, the required time is divided by \textit{n}.\\
The memory and complexity requirements for the process are heavily dependent on the specific block cipher used. Some advanced block ciphers can be complex and memory consuming. Other encryption algorithms are considered lightweight in terms of memory and complexity. However, the memory requirement is heavily reduced by the technique of key arbitration used in the proposed system. Using this self-key-generation technique reduces the memory requirement in addition to noticeable reduction in traffic between the KDA and the IoT device.\\
the size of the search space depends on the encryption key length and the number of IoT devices as well. With a key-length of \textit{u} bits, and \textit{n} IoT devices, each IoT device's sub-space has $\frac{2^u}{n}$. For older encryption standards like DES, where the key size is 56 bits, finding the key can be an easy task. With a total key search space of $2^{56}=7.205*10^{16}$, if the system employs one million devices, the sub-space for each device would be around 72057594038 keys. If in each second 1000 keys are used, a total of about 834 days are needed. If for the same setup 100 million devices are used, the key will be found within 8 days.\\
For complex and modern encryption algorithms like AES, key sizes and time to decrypt increase rapidly. if a 512-bit key is used with and arrangement similar to the DES example earlier, the time needed will be $1.551*10^{138}$ days, using 100 million devices, and processing 1000 keys/second.\\
Employing IoT devices does not necessary involve hijacking control of these devices. When IoT devices become part of every house, we might get to a point where people can rent out their IoT devices for similar uses. This can also involve renting only the free time of these IoT devices. There can be in the future services for device processing power rent out, similar to the idea of AirBnB, where people would enlist their devices and their availability time.
\section{Conclusion and Future Work}
The proposed conceptual design shows that IoT devices can be used in a collective manner to search for the encryption key of a block cipher in a brute-force attack. However, when the block cipher used has a large key size with complex mathematical operations, the proposed system can be slow, unless it employs a massively large number of IoT devices.\\
In our future work, we will be working on implementing the proposed design on a small scale to prove the concept proposed here. Another dimension of future work is applying the proposed system to asymmetric ciphers.

\bibliographystyle{IEEEtran}
\bibliography{references}
\end{document}